\documentclass[12pt]{iopart}
\usepackage{graphicx}% Include figure files
\usepackage{bm}% bold math
\usepackage{hyperref}% add hypertext capabilities

\begin{document}

\title[Nolocality and matrix product states]{Bell-CHSH function approach to quantum phase transitions in matrix product systems}

\author{Zhao-Yu Sun$^1$, Hai-Lin Huang$^1$ and Bo Wang$^2$ }

\address{$^1$ School of Electrical and Electronic Engineering, Wuhan Polytechnic University, Wuhan 430000, China}
\address{$^2$ Department of Physics, Beijing Normal University, Beijing 100875, China}

\ead{sunzhaoyu2012@gmail.com}
\begin{abstract}
Recently, nonlocality and Bell inequalities have 
been used to investigate quantum phase transitions (QPTs) in low-dimensional quantum systems. Nonlocality
can be detected by the Bell-CHSH function (BCF). In this work,
we extend the study of BCF to the QPTs in matrix product systems (MPSs). In this kind of QPTs, the ground-state energy 
keeps analytical in the vicinity of the QPT points, and is usually called the MPS-QPTs.
For several typical models, our results show that BCF can signal the MPS-QPTs very well.
In addition, we find BCF can capture signal of QPTs in unentangled states and classical
states, for which other measures of quantum correlation (quantum entanglement and quantum discord) fail. 
Furthermore, we find that in these MPSs, there exists some kind of quantum
correlation which cannot be characterized by entanglement, or by nonlocality.

\end{abstract}

\submitto{\JPCM}
\maketitle

\section{Introduction}

Quantum phase transition (QPT) is a very interesting phenomenon in
many-body quantum systems.\cite{BOOK} Compared with the classical
phase transitions driven by thermodynamic fluctuation, QPTs occur
at zero temperature, thus the thermodynamic fluctuation
is absent. In fact, QPTs are driven by the so-called quantum fluctuation.
For a quantum system described by a Hamiltonian $\hat{H}(g)$
with $g$ the tuning parameter, the ground-state property (i.e. the ground-state energy) of
$\hat{H}(g)$ may show qualitative change at some point $g_{c}$,
then a QPT occurs. Most QPTs, such as various magnetization transitions
in spin models,\cite{BOOK,QE4} can be investigated by traditional order
parameters. It needs mention that, some exotic QPTs, such as the
topological QPT,\cite{Quantum_topological_QPT_at_microscopic_level}
cannot be described by local order parameters.

In the vicinity of the QPT point, long-range correlations would develop
in the ground state. Thus it is expected that quantum correlation
plays a central role in the QPTs.\cite{Quantum_correlations_topological_QPT,Quantum_topological_QPT_at_microscopic_level,Bell_inequalitiesQPTs_XXZ_model,BOOK,Classical_correlation_and_quantum_discord_in_critical_systems,Correlation_nonlocalityQPTS_several_systems,QE4}
Quantum entanglement is the most famous measure of quantum correlation.
For various models, it has been found that the entanglement is singular
 in the vicinity of the QPT points, which is usually related to the singularity of the ground-state energy at the QPTs.\cite{QE4,re_examin_CC-1}
Nonlocality is another aspect of quantum correlation, and can be
indicated by the violation of Bell inequalities, such as the famous
Bell-Clauser-Horne-Shimony-Holt (CHSH) inequality.\cite{Bell_inequalities_in_Heisenberg_spin_chains,Bell_inequalitiesQPTs_XXZ_model,Correlation_nonlocalityQPTS_several_systems,Nonlocality_entanglement_XY_model,Nonlocality_entanglement__qubit_systems,Bell_Bell_Inequalityies,Horodecki_BCF_twoQubitState}
For a long time, nonlocality and entanglement were regarded as identical.
In fact, for pure two-qubit states, it has been proved rigorously
by Gisin that the two are indeed equivalent.\cite{Gisin_Entanglement_is_Nonlocal_twoQubitPureState}
However, for mixed states, it's found that an entangled state may
not violate any Bell inequality, i.e., an entangled state is not necessarily
a non-local state.\cite{Werner_Bell_Ineaulity_mixed_state} Thus,
nonlocality and entanglement turn out to be two different aspects
of quantum correlation.

Quite recently, it has been found that the Bell-CHSH function (BCF),
which is associated with the Bell-CHSH inequality, can serve as a
useful QPT detector, even for topological QPT\cite{Quantum_topological_QPT_at_microscopic_level,BCF_Topological_QPT}
and Kosterlitz-Thouless QPT\cite{Bell_inequalitiesQPTs_XXZ_model}.
It should point out that a complete understanding about the features
of BCF in detecting QPTs has not yet been reached. 

In this work, we make a further step by considering an exotic type of QPTs 
occurring in the so-called matrix product states(MPSs).\cite{MPS_discord,MPS_entanglement,MPS_Fidelity,MPS_fidelity_entanglement,MPS_ladder,MPS_QPT,MPS_XYZ}
In this kind of QPTs, the ground-state
energy remains analytic in the entire parameter space, thus the situation
is different from traditional QPTs.\cite{MPS_QPT}
In order to distinguish them from traditional
QPTs, we usually call them MPS-QPTs.
The MPSs
have always provided a valuable test-bed for understanding the features
of various QPT detectors, including quantum entanglement,\cite{MPS_entanglement,MPS_fidelity_entanglement}
quantum discord\cite{MPS_discord} and quantum fidelity\cite{MPS_Fidelity,MPS_fidelity_entanglement}.
In this work, we investigate the ability of BCF to detect QPTs by considering MPS models. Firstly,
as we will show, several typical MPSs display clearly the features
of BCF in detecting QPTs.
Secondly, MPSs help us understand the role of quantum correlation, quantum entanglement, and nonlocality in QPTs.

%This work is organized as follows. In Sec. II, we introduce the formula
%for calculating BCF, the entanglement concurrence and the discord. In Sec.
%III, firstly, the concept of MPS-QPT is introduced, and then we consider 
%bipartite nonlocality in several typical MPS models. A summary is given in Sec. IV.

\section{Bell-CHSH inequality, entanglement concurrence, and quantum discord}

Bell-CHSH inequality is the simplest nontrivial Bell inequality. First,
let's define the CHSH operator as $\hat{B}=\hat{A}_{1}\otimes\hat{B}_{1}+\hat{A}_{1}\otimes\hat{B}_{2}+\hat{A}_{2}\otimes\hat{B}_{1}-\hat{A}_{2}\otimes\hat{B}_{2}$,
where $\hat{A}_{i}=\vec{a}_{i}\cdot\vec{\sigma}$ and $\hat{B}_{i}=\vec{b}_{i}\cdot\vec{\sigma}$,
with $\vec{a}_{i}$ and $\vec{b}_{i}$ unit vectors and $\vec{\sigma}=(\hat{\sigma}_{x},\hat{\sigma}_{y},\hat{\sigma}_{z})$.
Then for any realistic and local two-qubit state $\hat{\rho}_{2}$, the Bell-CHSH
inequality reads $\vert\langle\hat{B}\rangle\vert=\vert\textrm{Tr}(\hat{\rho}_{2}\hat{B})\vert\le2$.
$\vert\langle\hat{B}\rangle\vert$ depends upon the vectors $\vec{a}_{i}$
and $\vec{b}_{i}$, and one can optimize $\vert\langle\hat{B}\rangle\vert$
over all vectors $\vec{a}_{i}$ and $\vec{b}_{i}$ to get the maximum
value $B(\hat{\rho}_{2})=\textrm{max}_{\{\vec{a}_{i},\vec{b}_{i}\}}\vert\langle\hat{B}\rangle\vert$.\cite{Nonlocality_entanglement__qubit_systems,Bell_inequalities_in_Heisenberg_spin_chains,Bell_Bell_Inequalityies}
We will refer to $B(\hat{\rho}_{2})$ as the Bell-CHSH function (BCF) in
this work. For some state $\hat{\rho}_{2}$, if it turns out that
$B(\hat{\rho}_{2})>2$, we usually say that the Bell-CHSH inequality is
violated, which means that the state $\hat{\rho}_{2}$ cannot be described
by a realistic local theory, in other words, it is non-local.

Alternatively, it's found by Horodeckis that for any two-qubit state
\begin{equation}
\hat{\rho}_{2}=\left(\begin{array}{cccc}
x_{11} & o_{12} & o_{13} & x_{14}\\
o_{21} & x_{22} & x_{23} & o_{24}\\
o_{31} & x_{32} & x_{33} & o_{34}\\
x_{41} & o_{42} & o_{43} & x_{44}
\end{array}\right),\label{eq:general}
\end{equation}
BCF can be expressed by a closed analytical formula.\cite{Horodecki_BCF_twoQubitState}
For convenience, here we use two works $x$ and $o$ to denote the
elements of $\hat{\rho}_{2}$, 
and for all the models considered in this work, 
it holds that $o_{ij}=o$. 
To calculate BCF, one first defines
a $3\times3$ matrix $\hat{L}$ as $L_{ij}(\hat{\rho}_{2})=\textrm{Tr}[\hat{\rho}_{2}\cdot\hat{\sigma}_{i}\otimes\hat{\sigma}_{j}]$,
with $\{\hat{\sigma}_{1},\hat{\sigma}_{2},\hat{\sigma}_{3}\}$ the
Pauli matrices. Then the BCF is given by $B(\hat{\rho}_{2})=2\sqrt{u+v}$,
with $u$ and $v$ the two largest eigenvalues of the symmetric matrix
$\hat{L}^{T}\hat{L}$.

In order to understand the features of nonlocality, we will compare it to
two closely related measures for bipartite correlation, i.e., entanglement concurrence\cite{def_CC}
and quantum discord\cite{Quantum_Discord_A_Measure_of_the_Quantumness_of_Correlations,Quantum_discord_for_two-qubit_X_states}.
Concurrence describes
the entanglement between two spins. Let's denote $\tilde{\rho}_{2}$
as the spin-flipped matrix for the two-qubit density matrix $\hat{\rho}_{2}$,
i.e., $\tilde{\rho}_{2}=\hat{\sigma}_{y}\otimes\hat{\sigma}_{y}\hat{\rho}_{2}^{*}\hat{\sigma}_{y}\otimes\hat{\sigma}_{y}$,
then the concurrence is given by $C=\textrm{max}\{0,\mu_{1}-\mu_{2}-\mu_{3}-\mu_{4}\}$,
where $\mu_{i}$ are the square roots of the eigenvalues of $\hat{\rho}_{2}\tilde{\rho}_{2}$
in decreasing order. For separable states (in other words, unentangled states), the concurrence would vanish,
and for maximum entangled states, the concurrence is $1$.

Nonlocality and entanglement are two aspects of quantum correlation. Recently, quantum discord is
proposed 
to characterize all the quantum correlation present in the system.\cite{Quantum_Discord_A_Measure_of_the_Quantumness_of_Correlations,Quantum_discord_for_two-qubit_X_states}
Its definition is based on two quantum versions of the classical correlation.
For a classical system $AB$ composed of two subsystems $A$ and $B$,
the total correlation can be expressed as $I_{A,B}=H_{{A}}+H_{{B}}-H_{{AB}}$,
or alternatively, $J_{A,B}=H_{{A}}-H_{{A}|{B}}$, with $H_{{A}}$,
$H_{{B}}$ and $H_{{AB}}$ the Shannon entropy, and $H_{{A}|{B}}$
the conditional entropy. $I_{A,B}$ and $J_{A,B}$ are equal to each
other, however, their quantum versions are found to be non-equivalent
from each other, and the difference is used to define the quantum discord. 
By replacing the Shannon entropy and the conditional
entropy with the von Neumann entropy and quantum conditional entropy,
respectively, $I_{A,B}$ becomes the quantum mutual information $\mathcal{I}(\hat{\rho}_{{AB}})=S(\hat{\rho}_{{A}})+S(\hat{\rho}_{{B}})-S(\hat{\rho}_{{AB}}),$
and the quantum extension of $J_{A,B}$ turns out to be the classical
correlation $\mathcal{J}(\hat{\rho}_{{AB}})=\textrm{max}_{\{\hat{B}_{k}\}}\{S(\hat{\rho}_{{A}})-S(\hat{\rho}|\{\hat{B}_{k}\})\}$,
where $\{\hat{B}_{k}\}$ is just a complete set of projectors.\cite{Quantum_discord_for_two-qubit_X_states}
Finally, discord is just defined as the difference between $\mathcal{I}(\hat{\rho}_{{AB}})$
and $\mathcal{J}(\hat{\rho}_{{AB}})$, i.e., $D(\hat{\rho}_{{AB}})=\mathcal{I}(\hat{\rho}_{{AB}})-\mathcal{J}(\hat{\rho}_{{AB}}).$
For a state containing quantum correlation, discord is generally non-zero,
while for classical states, $\mathcal{I}(\hat{\rho}_{{AB}})$ and
$\mathcal{J}(\hat{\rho}_{{AB}})$ would just reduce to $I_{A,B}$
and $J_{A,B}$, respectively, thus the discord vanishes.

From the above descriptions, one can see that the entanglement concurrence 
and quantum discord would simply be zero in separable states and classical
states, respectively.\cite{SunEpl} As we will show in the next section,
BCF can capture the signal of QPTs in these two situations.

\section{QPTs in MPSs}

In this section, we firstly give a brief introduction to MPSs, 
then we investigate the BCF and nonlocality at MPS-QPTs in several typical models.

An MPS containing $N$ sites (or cells) is defined in the following 
%Let's consider one-dimensional quantum systems with $N$ sites, and the degree of freedom
%for every site is $d$. It's found that for many systems, the ground states can be exactly
%expressed in the following 
matrix product form\cite{MPS_QPT} 
\begin{equation}
    \vert\psi(g)\rangle=\sum_{i_{1},...,i_{N}=1}^{d}\textrm{Tr}(\hat{A}_{i_{1}}...\hat{A}_{i_{N}})\vert i_{1},...,i_{N}\rangle,\label{eq:MPS}
\end{equation}
where $\hat{A}_{i}$ are $D\times D$ matrices, $j=1,...,N$ labels the sites, and $i_j=1,...,d$ denoting the 
degree of freedom for site $j$. The matrices $\hat{A}_{i}:=\hat{A}_{i}(g)$
depend on the parameter $g$.
%for every site is  . 
%For all the models
%considered in this work, the value of $D$ is just $2$.
%One can see that for a given
%set of matrices $\{\hat{A}_{i}\}$, the wavefunction in Eq. (\ref{eq:MPS})
%is determined completely. 
%If the ground-state wavefunction of a model can be analytically expressed as 
%the above matrix product form, we usually call the model a matrix product system(MPS).
%详细说明什么是MPS-QPT，它跟一般的有什么不同。

Usually, for low-dimensional quantum systems, it is difficult to express the ground-state wavefunction
exactly in an explicit form. However, for an MPS $\vert\psi(g)\rangle$, 
one can construct a parent Hamiltonian $\hat{H}(g)$, 
which guarantees the state $\vert\psi(g)\rangle$ be the ground state of $\hat{H}(g)$.
More importantly, in the thermodynamic limit $N\rightarrow\infty$, as the change of $g$,
the ground state $\vert\psi(g)\rangle$ may undergo a novel type of transition,
such that local observables are singular and the correlation length is divergent, 
with the ground-state energy keeping 
analytic (In traditional QPTs, the ground-state energy would be singular.)\cite{MPS_QPT}.
We usually say that the system undergoes an MPS-QPT.

For a given MPS,
the reduced density matrix of any subsystem in the system can be obtained
with the help of transfer matrix technique.\cite{MPS_XYZ,MPS_ladder,MPS_QPT}
First, let's define the transfer matrix $\hat{E}$ as 
\begin{equation}
\hat{E}=\sum_{i=1}^{d}\hat{A}_{i}^{*}\otimes\hat{A}_{i},\label{eq:E}
\end{equation}
 then the reduced density matrix of $k$ adjacent sites is given by
\begin{equation}
\rho_{i_{1},...,i_{k},j_{1},...,j_{k}}(N)=\frac{\textrm{Tr}[(\hat{A}_{i_{1}}^{*}...\hat{A}_{i_{k}}^{*}\otimes\hat{A}_{j_{1}}...\hat{A}_{j_{k}})]}{\textrm{Tr}(\hat{E}^{N})}.\label{eq:rho_general}
\end{equation}

One can further prove that, for two-site correlation, the correlation length is given by
$\xi=\frac{1}{\ln(\lambda_{1}/\lambda_{2})}$,
where $\lambda_1$ and $\lambda_2$ are 
the first and the second largest eigenvalue of the transfer matrix $\hat{E}$. 
Any level crossing between 
$\lambda_1$ and $\lambda_2$ indicates a divergent correlation length, in other words, an MPS-QPT.\cite{MPS_QPT}

In this work, we only deal with bipartite correlations.
For several typical MPS models, we use Eq. (\ref{eq:rho_general})
to obtain the reduced density matrix for the concerned two-site subsystem of the models.
Then we determine the BCF,
and research the behavior of BCF, concurrence and discord in the MPS-QPTs.

\subsection{Spin ladder with four-body interaction}

As the first example, we consider an
MPS with $\hat{A}_{1,2}=\left(\begin{array}{cc}
a & 0\\
0 & a
\end{array}\right)$, $\hat{A}_{3}=\left(\begin{array}{cc}
0 & g\\
g & 0
\end{array}\right)$, $\hat{A}_{4}=\left(\begin{array}{cc}
0 & 0\\
1 & 0
\end{array}\right)$.\cite{MPS_ladder} Its 
parent Hamiltonian describes a spin $s=\frac{1}{2}$ ladder model with SO(2) symmetry.
Every rung of the ladder contains two spins with $d=4$, thus four matrices are used to define the MPS.
The system contains two-body bond interactions and four-body plaquette interactions.\cite{MPS_ladder}
As the Hamiltonian is too long, we would not show it in this work.
It has been found that the ladder undergoes an MPS-QPT at $g=0$, where the spin-spin correlation
function of the ladder shows a singularity 
and the largest two eigenvalues of the transfer matrix have a level crossing.
%说明白，此处本征值交叠。所以确实是相变点。

\begin{figure}
\includegraphics{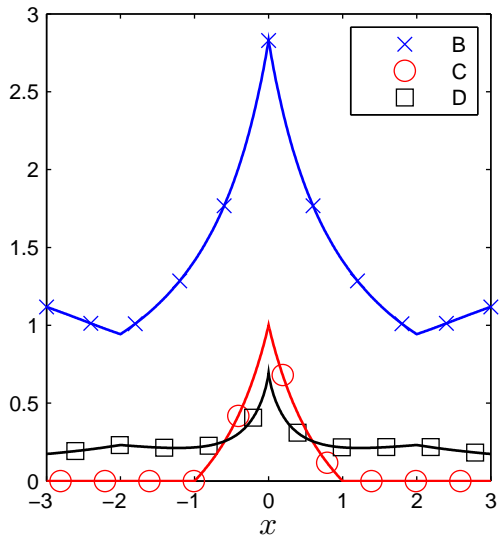}\caption{\label{fig:Ladder} (Color Online) The BCF (B), concurrence (C), and discord (D) as a function of $x=\frac{g}{2a^{2}}$
for the rung of the ladder. The concurrence 
has been investigated in Ref. \cite{MPS_ladder} and is shown
here just for comparison purpose. }
\end{figure}

Now we try to use quantum correlations to find the signal for the MPS-QPT.
Let's just consider a single rung in the ladder.  The elements of the 
reduced density matrix $\hat{\rho}^{(\textrm{rung})}_2$ turn out to be
\begin{equation}
\begin{array}{l}
\{x_{11},x_{44}\}=\vert x\vert\\
\{x_{22},x_{23},x_{32},x_{33}\}=1\\
\{o,x_{14,}x_{41}\}=0
\end{array},
\end{equation}  
where $x:=\frac{g}{2a^{2}}$. We have determined BCF on the
rung and shown it in Fig. \ref{fig:Ladder}. The BCF shows a
singularity at the MPS-QPT point $g=0$, thus it can be used to detect the MPS-QPT in this model.

In addition, the first-order derivative of BCF is 
discontinuous at $x=\pm2$. Detailed analysis shows that the
singular points at $x=\pm2$ are due to the mathematical definition of BCF, 
rather than the singularity in $\hat{\rho}^{(\textrm{rung})}_2$.
Explicitly, the non-physical singularity is induced by the max function in
the definition of BCF. In order to calculate BCF, one has to find the two largest eigenvalues of the symmetric matrix
$\hat{L}^{T}\hat{L}$. In this procedure, a mathematical singularity may emerge. In fact, a max/min function is also
involved in the definition of the discord/concurrence. As a result, the concurrence and discord can also show a
non-physical singularity, just as shown in Fig. 1.

Now let's discuss the feature of quantum correlation in the rung.
As indicated by the discord in Fig. \ref{fig:Ladder}, quantum correlation exists for any finite $x$.
In the vicinity of the MPS-QPT point, that is, for $\vert x\vert<0.41$,
it's found that the concurrence is non-zero and $B>2$, thus the quantum
correlation is in the form of both entanglement and nonlocality. While
for $0.41<\vert x\vert<1$, the concurrence is non-zero and $B<2$,
thus the quantum correlation is in the form of entanglement without nonlocality. For $\vert x\vert>1$, it is present neither in the form of entanglement nor in the form of nonlocality.
It shows clearly that quantum correlation can be manifested by various forms.

\subsection{XYZ interaction model}

\begin{figure}
\includegraphics{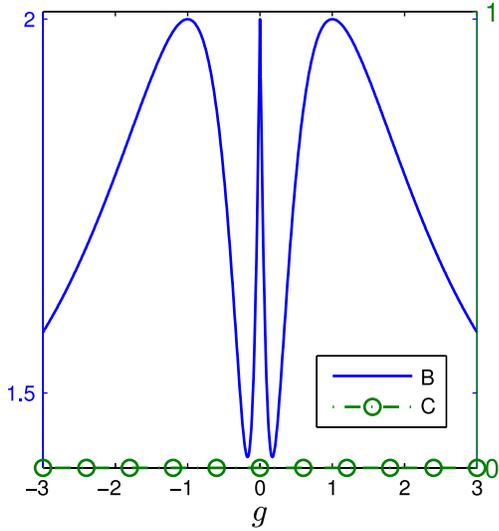}\caption{\label{fig:XYZ} (Color Online) The BCF (B) as a function of $g$ for any two spins
in the XYZ interaction model. The concurrence (C) has been investigated
in Ref. \cite{MPS_XYZ} and is shown here just for comparison purpose.}
\end{figure}

Let's consider an MPS with $\hat{A}_{1}=\left(\begin{array}{cc}
1 & g\\
1 & 1
\end{array}\right)$ and $\hat{A}_{2}=\left(\begin{array}{cc}
1 & -g\\
-1 & 1
\end{array}\right)$. 
After the standard procedure, one can construct its parent Hamiltonian as \cite{MPS_XYZ} 
\begin{equation}
\hat{H}=\sum_{i=1}^{N}J_{x}\hat{\sigma}_{x}^{i}\hat{\sigma}_{x}^{i+1}+J_{y}\hat{\sigma}_{y}^{i}\hat{\sigma}_{y}^{i+1}+J_{z}\hat{\sigma}_{z}^{i}\hat{\sigma}_{z}^{i+1}-B\hat{\sigma}_{z}^{x},
\end{equation}
with $J_{x}=-J+\frac{1}{2}(1+g^{2})$,
$J_{y}=-J+g$, $J_{z}=-J-g$ and $B=1-g^{2}$. It is just an XYZ interaction chain. 
It has been proved that the system has an MPS-QPT at $g=0$.\cite{MPS_XYZ}

Firstly, we consider the correlation between two nearest-neighboring spins $i$ and $i+1$. For $g>0$, the 
corresponding reduced density matrix $\hat{\rho}_{2}^{(i,i+1)}$ is given by 
\begin{equation}
\begin{array}{l}
\{x_{11},x_{44}\}=g^{2}+6g+1\\
\{x_{22},x_{23},x_{32},x_{33},x_{14},x_{41}\}=(g-1)^{2}\\
o=1-g^{2}
\end{array},
\end{equation}  
 while for $g>0$, one finds that 
\begin{equation}
\begin{array}{l}
\{x_{14},x_{41}\}=g^{2}+6g+1\\
\{x_{22},x_{23},x_{32},x_{33},x_{11},x_{44}\}=(g-1)^{2}\\
o=1-g^{2}
\end{array}.
\end{equation}  

Previous studies show that $\hat{\rho}_{2}^{(i,i+1)}$ is separable for any $g$, 
thus the bipartite entanglement between $i$ and $i+1$ vanishes
and cannot detect the MPS-QPT of the system.\cite{MPS_XYZ,MPS_discord}
From Fig. \ref{fig:XYZ} we find that BCF is generally non-zero in the whole parameter
space and shows a singularity at the MPS-QPT point $g=0$. 

In fact, for any two-site subsystems of the model, the
reduced density matrix $\hat{\rho}_{2}^{(i,i+r)}$ does not
depend on the distance $r$ at all.\cite{MPS_XYZ} As a result,
the concurrence for any two-site subsystem is zero,  
thus bipartite entanglement cannot detect the QPT of the system while the BCF can do the job very well.
This example shows clearly that BCF can detect QPTs in separable states while entanglement fails.

We observe that the Bell-CHSH inequality is not violated in the MPS-QPT in this XYZ interaction model.
We'd like to mention that the quantum correlation indeed
exists in the QPT region, indicated by the discord.\cite{MPS_discord} It is interesting that the
quantum correlation present in $\hat{\rho}_{2}^{(i,i+r)}$ is neither in
the form of entanglement nor in the form of nonlocality.

\subsection{Three-body interaction model}

We consider an MPS with
$\hat{A}_{1}=\left(\begin{array}{cc}
0 & 0\\
1 & 1
\end{array}\right)$ and $\hat{A}_{2}=\left(\begin{array}{cc}
1 & g\\
0 & 0
\end{array}\right)$. Its parent Hamiltonian describes a three-body interaction model as follows \cite{MPS_QPT} 
\begin{equation}
\hat{H}=\sum_{i=1}^{N}J_{3}\hat{\sigma}_{z}^{i}\hat{\sigma}_{x}^{i+1}\hat{\sigma}_{z}^{i+2}+J_{z}\hat{\sigma}_{z}^{i}\hat{\sigma}_{z}^{i+1}-B\hat{\sigma}_{i}^{x},
\end{equation}
with $J_{3}=(g-1)^{2}$, $J_{z}=2(g^{2}-1)$,
and $B=(1+g)^{2}$. The system undergoes an MPS-QPT at $g=0$.\cite{MPS_QPT}

\begin{figure}
\includegraphics{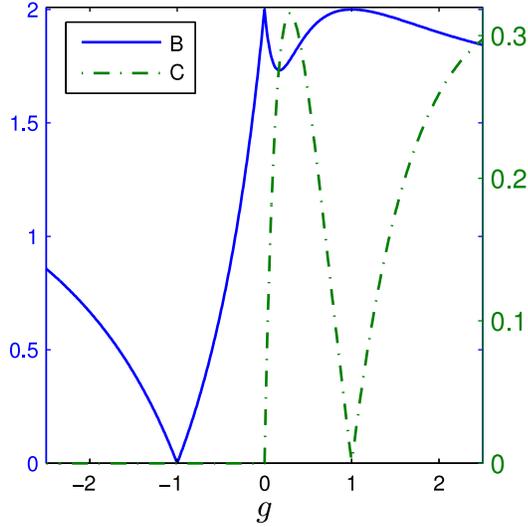}\caption{\label{fig:Three_1} (Color Online) The BCF (B) as a function of $g$ for nearest-neighboring
two sites in the three-body interaction model. The concurrence (C)
has been investigated in Ref. \cite{MPS_discord} and is shown just
for comparison purpose.}
\end{figure}

We consider the two-qubit states $\hat{\rho}_{2}^{(i,i+r)}$ in the
chain with different $r$. When $r=1$, the reduced density matrix for $g>0$ is given by 
\begin{equation}
\begin{array}{lc}
\{x_{11},x_{44}\}=\frac{g+1}{2}, & \{x_{22},x_{33}\}=\frac{g^{2}+g}{2}\\
\{x_{23},x_{32}\}=\frac{2g^{2}}{g+1}, & \{x_{14},x_{41}\}=\frac{2g}{g+1}\\
o=g
\end{array},
\end{equation} 
while for $g<0$, $\hat{\rho}_{2}^{(i,i+1)}$ is a diagonal matrix,
with the diagonal entries given by 
\begin{equation}
\begin{array}{cc}
\{x_{11},x_{44}\}=1, & \{x_{22},x_{33}\}=-g\end{array},
\end{equation} 
which denotes a classical state. The BCF is shown
in Fig. \ref{fig:Three_1}. BCF shows a singularity at $g=0$,
thus it can be used to detect the MPS-QPT of the model. In addition, BCF is singular
at $g=-1$. Detailed analysis shows that this singular point results
from the mathematical definition of BCF, rather than a
transition in $\vert\psi(g)\rangle$. For $g>0$, we observe that $\hat{\rho}_{2}^{(i,i+1)}$ never
violate the Bell inequality, despite being entangled. In other words,
the quantum correlation in nearest-neighboring sites is in the form
of entanglement, rather than nonlocality.

Next, we consider two sites $i$ and $i+r$ with $r\ge2$. For $g>0$,
the elements of the reduced density matrix $\hat{\rho}_{2}^{(i,i+r)}$
are given by 
\begin{equation}
\begin{array}{l}
\{x_{11},x_{44}\}=1+(\frac{1-g}{1+g})^{r}\\
\{x_{22},x_{33}\}=1-(\frac{1-g}{1+g})^{r}\\
\{x_{14},x_{23},x_{32},x_{41}\}=\frac{16g^{2}}{(1+g)^{4}}\\
o=\frac{4g}{(1+g)^{2}}
\end{array},
\end{equation}  
while for $g<0$, $\hat{\rho}_{2}^{(i,i+r)}$ is reduced to a diagonal
matrix, with the diagonal elements given by 
\begin{equation}
\begin{array}{cc}
\{x_{11},x_{44}\}=1+(\frac{1+g}{1-g})^{r}, & \{x_{22},x_{33}\}=1-(\frac{1+g}{1-g})^{r}\end{array}.
\end{equation}

\begin{figure}
\includegraphics{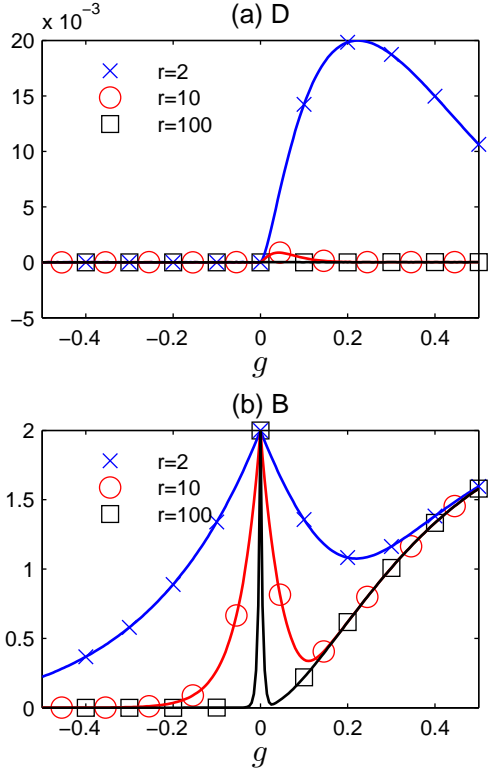}\caption{\label{fig:Three_2} (Color Online) The quantum discord (D) and the BCF (B) as a
function of $g$ for two sites $i$ and $i+r$ in the three-body interaction
model.}
\end{figure}

We numerically found that the concurrence vanishes in the whole parameter
space for $r\ge2$, which means that $\hat{\rho}_{2}^{(i,i+r)}$ is
separable. Previous study shows that quantum discord may be able to
capture the signal of QPT in separable states\cite{SunEpl,MPS_discord}.
We have calculated the discord for different $r$ and the result
is shown in Fig. \ref{fig:Three_2}(a). One can see that the value of discord is very
small even for $r=2$. As the increase of the distance $r$, the discord
decreases rapidly. Finally, for a large $r$, the two-qubit state
$\hat{\rho}_{2}^{(i,i+r)}$ would become a classical state without
any quantum correlation, thus neither the concurrence nor the discord
can signal the MPS-QPT of the system. Then let's study the BCF for $\hat{\rho}_{2}^{(i,i+r)}$.  
From Fig. \ref{fig:Three_2}(b) one sees clearly that BCF shows a singularity
at the QPT point $g=0$ for any finite $r$. Thus, BCF is able to
capture the singularity in classical states, for which both
the concurrence and the discord fail. In addition, our results show
that the Bell-CHSH inequality is never violated, thus the quantum
correlation between the non-nearest-neighbor spins, if exists, is neither in the form
of entanglement, nor in the form of nonlocality.

\section{Summaries and discussions}

%BCF捕捉QPT信号的一般研究。
In this work, for several typical models,
we find that BCF can be used to detect the MPS-QPTs very well.
The underlying mechanism is as follows. 
Our discussion applies to MPS-QPTs and traditional QPTs. 
When a QPT occur,
some local observables, such as spin-spin correlation functions, would be singular at the QPT point.\cite{Note}
These correlation functions can be used to construct the reduced density matrix $\hat{\rho}_2(g)$.\cite{PRB_Sun_2008}
As a result, the QPT and the singularity in $\hat{\rho}_2(g)$ are closely related to each other.
The BCF, concurrence and discord, defined based upon $\hat{\rho}_2(g)$,
may capture the singularity in $\hat{\rho}_2(g)$, thus, all the three quantities 
can be used to detect the QPTs.

%BCF的优势，详述。相对而言，QE和QD都有不足。
In Ref. \cite{Bell_inequalitiesQPTs_XXZ_model} a general argument for why BCF should be as good as entanglement 
to signal QPTs has already been made. 
Now let's discuss the advantage of BCF in detecting QPTs.  
BCF can capture singularity in unentangled states and classical
states, for which entanglement and discord fails, respectively. 
Quantum discord is defined on
the classical-quantum paradigm from a measurement perspective, thus it captures all the quantumness of correlation in the system. 
For classical states, discord is zero thus loses it function as a QPT
indicator. This is just the situation for $\hat{\rho}^{(i,i+r)}_2$ with $r>1$ in the three-body interaction model. 
On the other hand, quantum entanglement is defined on the separability-entanglement paradigm.
For quantum separable states, it is zero thus would lose the signal of
QPT in these states. This is just what happens in any two-site subsystem in the XYZ model.
BCF is used to detect the nonlocality of a state. For
non-local states, it would be larger than $2$, while for most local
states, it does not vanish. In extreme cases, even if the state is a
classical state thus $\hat{\rho}_{2}$ is diagonal, the BCF would
turn out to be $B=2\vert x_{11}+x_{44}-x_{22}-x_{33}\vert$, which is still non-zero
unless $x_{11}+x_{44}=x_{22}+x_{33}$. As a result, BCF can detect
the singularity in various density matrices, including separable/entangled states,
local/non-local states, and classical/quantum states.

%事实上，不只是X states。
%
%The above advantage for BCF is just based upon Eq. (\ref{eq:B}), 
%thus it holds for any two-qubit density matrix which can be expressed 
%in the form of Eq. (\ref{eq:XP}).
%As we have discussed in Sec. II, for spin $s=1/2$ systems with parity invariance or spin-flip symmetry, 
%the reduced density matrix for a two-site subsystem should take the form of Eq. (\ref{eq:XP}). 
%We conclude that, for any spin $s=1/2$ system with parity invariance or spin-flip symmetry, 
%the two-site BCF 
%can capture the signal of QPTs even if the quantum entanglement or the quantum discord fail.

%BCF的不足。相对而言，QE和QD也有这种不足。
Now we clarify the drawback of BCF. First of all, for a general
state, the calculation of BCF is difficult, which greatly limits its
application. In addition, as shown in $x=\pm2$ of Fig. \ref{fig:Ladder} and $g=-1$ of Fig. \ref{fig:Three_1}, 
the mathematical definition of BCF 
can introduce non-physical singularity, which is not related to the singularity in $\hat{\rho}(g)$.
As a result, the singularity of BCF just can be used to detect, rather than to determine, an MPS-QPT.
However, we'd like to mention that both the concurrence and the discord
have similar disadvantage (see Fig. \ref{fig:Ladder} as examples),
which has already been discussed in some other studies.\cite{SunEpl,re_examin_CC-1}  

Let's discuss the form of quantum correlation in the MPS-QPTs of
these models. In this work, we have only dealt with bipartite correlations. 
We use the discord to identify the existence of quantum correlation, 
and then describe the nature of quantum correlation through the analysis of entanglement and nonlocality.
In the vicinity of MPS-QPT point in
the ladder model, the quantum correlation in $\hat{\rho}^{(\textrm{rung})}_2$ is in the form
of both entanglement and nonlocality. In the three-body interaction
model, for $\hat{\rho}_{2}^{(i,i+1)}$, the quantum correlation
is in the form of entanglement without nonlocality, while for $\hat{\rho}_{2}^{(i,i+r)}$ with $r>1$, 
the quantum correlation, if exist, is neither in the form of
entanglement nor nonlocality. In the XYZ interaction model, for
any two spins in the chain, the quantum correlation is neither in
the form of entanglement nor nonlocality. 
From one hand, our results
show that when MPS-QPTs occur, the two-site quantum correlation can show very rich nature 
combined with entanglement and nonlocality. From another
hand, it reveals that entanglement and nonlocality are not the only aspects of quantum
correlation, and there exists some kind of quantum
correlation which cannot be characterized by entanglement, nor by nonlocality.

Finally, as we have shown in this work, the Bell-CHSH inequality is violated just in the QPT of the four-body interaction ladder model.
In fact, as far as we know, in all the previous works,
\cite{BCF_Topological_QPT,Bell_inequalitiesQPTs_XXZ_model,Correlation_nonlocalityQPTS_several_systems} 
when QPTs occur in infinite models, the density matrices of two-qubit subsystems never violate the Bell-CHSH inequality.
Thus, the ladder model reported in this work may be the first one to present such a behavior.
In QPTs in many-body systems, as an important aspect of quantum correlation, nonlocality should play a central role,
however, it turns out that bipartite nonlocality is not a common form of quantum correlation present in these one-dimensional systems.

We'd like to mention that, for many-body systems, which are naturally multipartite, 
it would be more natural for quantum correlation to present in the form of 
multipartite nonlocality.\cite{multi_BFV, Multipartite_nonlocality_without_entanglement_in_many_dimensions, Multipartite_nonlocal_quantum_correlations_resistant_to_imperfections}
It has been found that Bell inequalities can be used to test multipartite nonlocality.\cite{Testing_for_Multipartite_Quantum_Nonlocality_Using_Functional_Bell_Inequalities}  
In addition, effective approaches to detect or even quantify multipartite nonlocality have been proposed.\cite{Quantifying_Multipartite_Nonlocality,
Detecting_Genuine_Multipartite_Quantum_Nonlocality_A_Simple_Approach_and_Generalization_to_Arbitrary_Dimensions}

However, 
the relevance of multipartite nonlocality in QPTs remains unknown.
As bipartite nonlocality is not favored at the QPT points in many one-dimensional models, 
it would be interesting to clarify whether multipartite nonlocality is significant in QPTs.
Considering the simple product form of MPSs, we believe MPS models would be very useful to study this important issue.

After finishing this manuscript, we become aware of a related work \cite{Important} by Oliveira et. al.
The authors have proposed a general explanation for why the Bell inequality is not violated
in most translation invariant systems. Then they have shown that the inequality can be violated 
for models with translation symmetry breaking. 
Our results of the spin ladder model suggest that, for two spins located in a unit-cell of a complex lattice, 
the Bell inequality can still be violated in
a translation invariant system.

\section*{Acknowledgment}

The research was supported by the National Natural Science Foundation of China (No. 11204223). 
This work was also supported by the Talent Scientific Research Foundation
of WHPU (No. 2012RZ09 and 2011RZ15), and China Postdoctoral Science
Foundation (No. 2012M510342).


\begin{thebibliography}{10}
\bibitem{Note} In some models, the transition in $\vert\psi(g)\rangle$ are not inherited 
by any two-site correlation. In these cases, all the measures of two-site quantum correlation,
including BCF, concurrence and discord, would not be able to detect the QPTs, and we have to 
adopt larger subsystems to investigate the QPTs. In this work, our discussions 
are limited to models in which the two-site correlations are able to inherite the transition of $\vert\psi(g)\rangle$.
\end{thebibliography}
\end{document}